# Overcoming scattering in high-cell-density tomographic volumetric bioprinting using computational light optimization

*Qianyi Zhang*, Felix Wechsler, Viola Sgarminato, Christophe Moser, Riccardo Rizzo**

Q. Zhang, F. Wechsler, Viola Sgarminato, C. Moser, R. Rizzo

Laboratory of Applied Photonics Devices, School of Engineering, École Polytechnique Fédérale de Lausanne, CH-1015, Lausanne, Switzerland

Corresponding authors: qianyi.zhang.scholar@gmail.com, riccardo.rizzo@epfl.ch

V. Sgarminato

Dipartimento di Ingegneria Meccanica e Aerospaziale (DIMEAS), Politecnico di Torino,

10129, Turin, Italy

## Abstract

Tomographic volumetric additive manufacturing has emerged as a transformative 3D printing technology for rapidly fabricating complex geometries. It offers significant advantages for bioprinting due to its contactless and short process time (a few tens of seconds). However, the presence of high cell densities (>$10^7$ cells $mL^{-1}$) in bioresins introduces substantial light scattering, which degrades printing resolution and fidelity, hindering the fabrication of biologically relevant microstructures such as vascular channels and cavities. To address this challenge, we utilize a computational patterning framework leveraging physically based inverse rendering to optimize light delivery in scattering environments. This method iteratively refines tomographic projections by simulating light-matter interactions in cell-laden hydrogels, enabling precise compensation for scattering effects. Experimental results demonstrate that our approach achieves 500 μm diameter vascular channels at $2\times10^7$ cells $mL^{-1}$. Furthermore, we integrate this computational method with refractive index matching strategies, reducing scattering artifacts by minimizing optical mismatch between cells and the hydrogel matrix, enabling printing at $4.1\times10^7$ cells $mL^{-1}$. The compatibility of these dual strategies enables unprecedented print fidelity in turbid bioresins. This advancement expands the scope of tomographic volumetric additive manufacturing for engineering functional tissues with intricate microarchitectures.

## 1. Introduction

Tomographic volumetric additive manufacturing (TVAM) enables rapid, whole-volume fabrication by projecting light patterns from various angles to accumulate a 3D dose distribution and cure the object in seconds[1], [2] (Figure 1a). This layer-free printing technique is especially attractive for bioprinting as it eliminates the need for support structures, facilitating the fabrication of complicated negative features such as vascular structures[3], [4], [5]. Moreover, its fast printing speed and contactless nature minimize cell stress and enable high-throughput testing applications such as drug screening.

However, to attain physiologically relevant tissue models, being able to bioprint at physiologically relevant high cell densities is fundamental[6], [7], [8]. To date, a central barrier to extending TVAM into biologically relevant, cell-dense hydrogels is optical scattering: cells and subcellular structures produce strong forward scattering that scrambles projected patterns and redistributes light dose, severely affecting the printing resolution and fidelity when printing at biologically relevant cell concentrations. Previous studies have demonstrated tomographic volumetric bioprinting at various cell densities and with different 3D structures[4], [9], [10], [11], [12]. Most of them are printed at cell densities below $5\times10^6$ cells $mL^{-1}$ and focus on relatively simple positive structures, with one study working at a density of $10^6$ cells $mL^{-1}$ using a meniscus-shaped construct[4], and another one at the same density to form a cartilage layer around a femoral head[12]. These studies indicate that it is challenging to print bulky constructs at high cell densities ($>10^7$ cells $mL^{-1}$). Printing fidelity also depends on the shape and the direction of the 3D constructs. In a scattering resin, negative features such as cavities and channels, commonly appearing in tissue-mimic constructs, are especially difficult to preserve in shape and function because scattered light increases the light dose in the cavities/channels and decreases the light dose contrast between the target and the surrounding regions, triggering photopolymerization in the channels and blocking these features. Moreover, so far, the bioprinting limit (e.g., cell density, printable structure, printing resolution, and fidelity) resulting from the optical scattering was largely based on empirical knowledge and qualitative studies, and remains to be explored thoroughly. Therefore, it is important to study the scattering effect on the printing fidelity and cell density limit of TVAM to guide the broader bioprinting applications of this technique.

Different scattering mitigation strategies have been proposed recently to address this issue in tomographic volumetric bioprinting. They can be divided broadly into material adjustment, optical intervention, and computational compensation. The material route utilizes high refractive-index agents such as iodixanol to reduce the refractive index mismatch between the subcellular structures and the hydrogels[13]. It effectively reduces the bulk turbidity of the bioresin but raises concerns about formulation complexity, altered polymerization kinetics, and biocompatibility. The optical approach uses self-healing beams, such as the Bessel beam, that penetrate deeper into the scattering media to encode the projected patterns[14], [15]. However, its effectiveness at high-cell-density resin is yet to be validated.

Computational route designs projected patterns to optimize the dose distribution for TVAM. The early approach of pattern generation used filtered backprojection with the negative values in the patterns, generated by the Fourier filter, clipped to zero[1], which can lead to severe artifacts in printing. Iterative approaches based on minimizing the difference between the reconstructed dose and the binary target dose have been proposed[2], [16]. Among them, object-space model optimization (OSMO) optimizes the model of the desired structure instead of the image set, resulting in a more desired dose reconstruction[16]. Some algorithms also model the absorption[2] and oxygen diffusion[17] in the printing and correct for them. To account for the geometry of the real optical system, Webber et al. introduced ray tracing to model the refraction at the air-vial and vial-resin interface[18]. Computational approaches have also been developed to model and correct for scattering. The deconvolution approach, based on filtered backprojection, treats scattering as a spatial convolution that dampens high frequencies, and a correction filter that amplifies the high-frequency components is constructed accordingly[3]. Its effectiveness was demonstrated at $4\times10^6$ cells $mL^{-1}$ using a construct with four channels (1.3 mm diameter). This approach shows promising results in isotropic scattering media, but can be challenging to implement in highly forward-scattering media, such as cell-loaded hydrogels, because it relies on the accurate recording of side views. Recently, an open-source physics-based inverse rendering algorithm (Dr.TVAM) has been proposed to model different physical effects, such as vial geometry, absorption, refraction, and scattering[19], [20]. It has been proven effective in an isotropic scattering medium, but has not been tested yet in strongly forward-scattering media such as cell-loaded hydrogels.

In this study, this inverse rendering algorithm is applied to tomographic volumetric bioprinting using hydrogels of high cell density ($2$–$4.1\times10^7$ cells $mL^{-1}$) in 5 mm inner diameter vials (Figure 1). Scattering properties of cell-loaded hydrogels are experimentally measured with a goniometric setup and serve as input to the algorithm, ensuring a more accurate simulation model for real bioprinting. In the algorithm (Figure 1b), patterns are generated and projected in a ray-tracing model with absorption, scattering properties, and geometry that match the experimental setup. Light dose distribution in the cell-loaded bioresin is then collected and compared with the target dose to calculate the loss function. After that, patterns for the next iteration are generated with gradient-based optimization. At high cell density, simulation and experimental results show that the targeted construct, a vascular model, fails to be printed by the pattern set generated without scattering taken into account (Figure 1c). In contrast, the construct can be correctly printed showing perfusable channels using the scattering-corrected pattern set (Figure 1d). Exploiting such physically based scattering correction, we demonstrate printing at a record-high cell concentration of $2\times10^{7}$ cells $mL^{-1}$. In addition, we demonstrate that combining the computational strategy with formulation adjustment (iodixanol) leads to even better performance, reaching successful prints with perfusable negative features at $4.1\times10^{7}$ cells $mL^{-1}$. These results extend the potential of tomographic volumetric bioprinting for the rapid biofabrication of centimeter-scale high-cell-density tissue models.

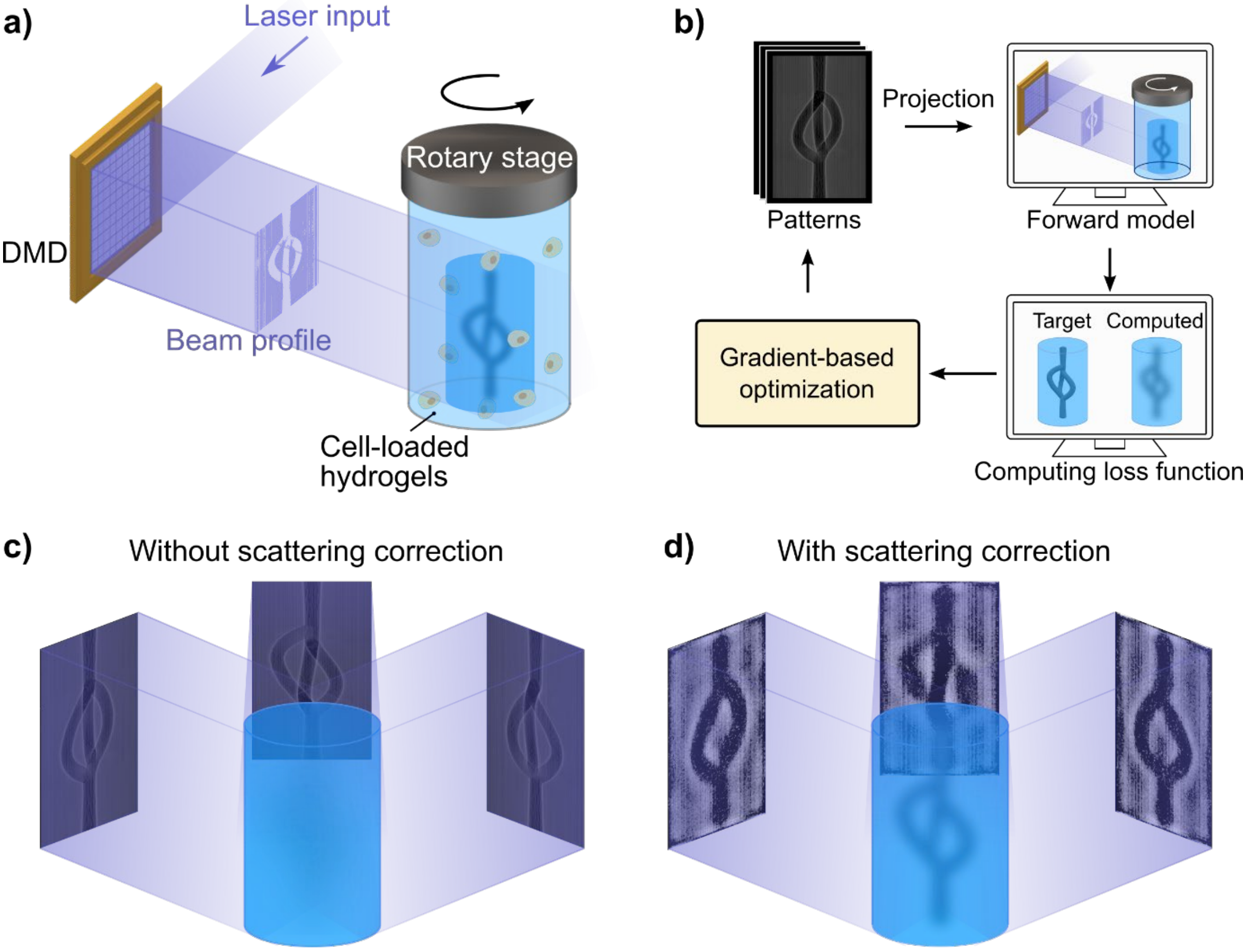

**Figure 1.** Working principle of overcoming scattering in tomographic volumetric bioprinting using computational light optimization. **a.** Schematic figure of tomographic volumetric bioprinting. **b.** Schematic diagram of the physics-based inverse rendering algorithm for generating scattering-corrected patterns. Schematic bioprinting results of projected patterns without (**c**) and with (**d**) scattering correction.

## 2. Results

### 2.1. Characterization of scattering properties

The scattering parameters used in the Monte Carlo ray-tracing simulation of Dr.TVAM are the scattering coefficient and the scattering phase function (Figure 2a). The scattering coefficient is defined as the probability per unit length that a photon is scattered in this medium, while the phase function describes the probability of a photon being diverted at various angles after a scattering event, as illustrated in Figure 2a. As the loss function for the pattern optimization is computed by comparing the simulation dose with the target dose (Figure 1b), it is critical to apply the scattering properties of the bioresin as the input to obtain an accurate simulation dose. A direct measurement of the scattering coefficient is usually conducted by measuring the attenuation of the transmitted light through a sample and calculating the scattering coefficient based on the Beer-Lambert law. A direct measurement of the phase function is usually conducted via a goniometer (a detector rotating around the sample) that measures scattered intensity versus polar angle. However, both measurements for cell-loaded hydrogels are quite challenging because of the forward scattering property of the media. The anisotropic factor $g$ (average cosine of scattering angle weighted by the phase function) of such media is usually higher than 0.96, meaning that most of the scattered light exhibits only a small deviation from the ballistic light, making it difficult to differentiate scattered light from the transmitted beam at small angles. For phase function measurement, most of the methods reported in the literature start at around 10° to avoid collecting the ballistic light, yet it is critical to collect data within 10° for a highly forward-scattering sample. A fixed focal lens placed after the sample has been proposed to push the lower measurement range from several degrees to less than one degree[21]. This configuration is adapted to our setup for phase function characterization (Figure 1d).

The scattering coefficient measurement is conducted by placing a sensor with a 1-mm-diameter pinhole at 0° to collect the ballistic light, as seen in Figure 2b. The samples are cell-loaded 7% gelatin methacryloyl (GelMA) hydrogels encapsulated in a cuvette, and the control group is pure 7% GelMA hydrogel in the same cuvette model. The transmission is calculated by dividing the transmitted light of the sample by that of the control group to only keep the attenuation resulting from the scattering of cells. Then the scattering coefficient is calculated according to the Beer-Lambert law. The scattering coefficient is measured at various cell densities relevant to bioprinting. The light path of cuvettes for scattering coefficient measurement is selected to obtain a higher attenuation in the sample while ensuring that the scattering signal at 0° is negligible compared to the ballistic signal. At 0.1–1×$10^6$ cells $mL^{-1}$, the light path is 10 mm; At 1–5×$10^6$ cells $mL^{-1}$, the light path is 2 mm; at 5–25×$10^6$ cells $mL^{-1}$, the light path is 0.5 mm. When switching cuvettes, the scattering coefficient at a certain cell density is measured using cuvettes of different light paths. The calculated scattering coefficients do not vary with the light path, indicating that the measurement is valid. The measured scattering coefficient is proportional to the cell density within the measured cell density range, as shown in Figure 2c, indicating the accuracy of this direct measurement method.

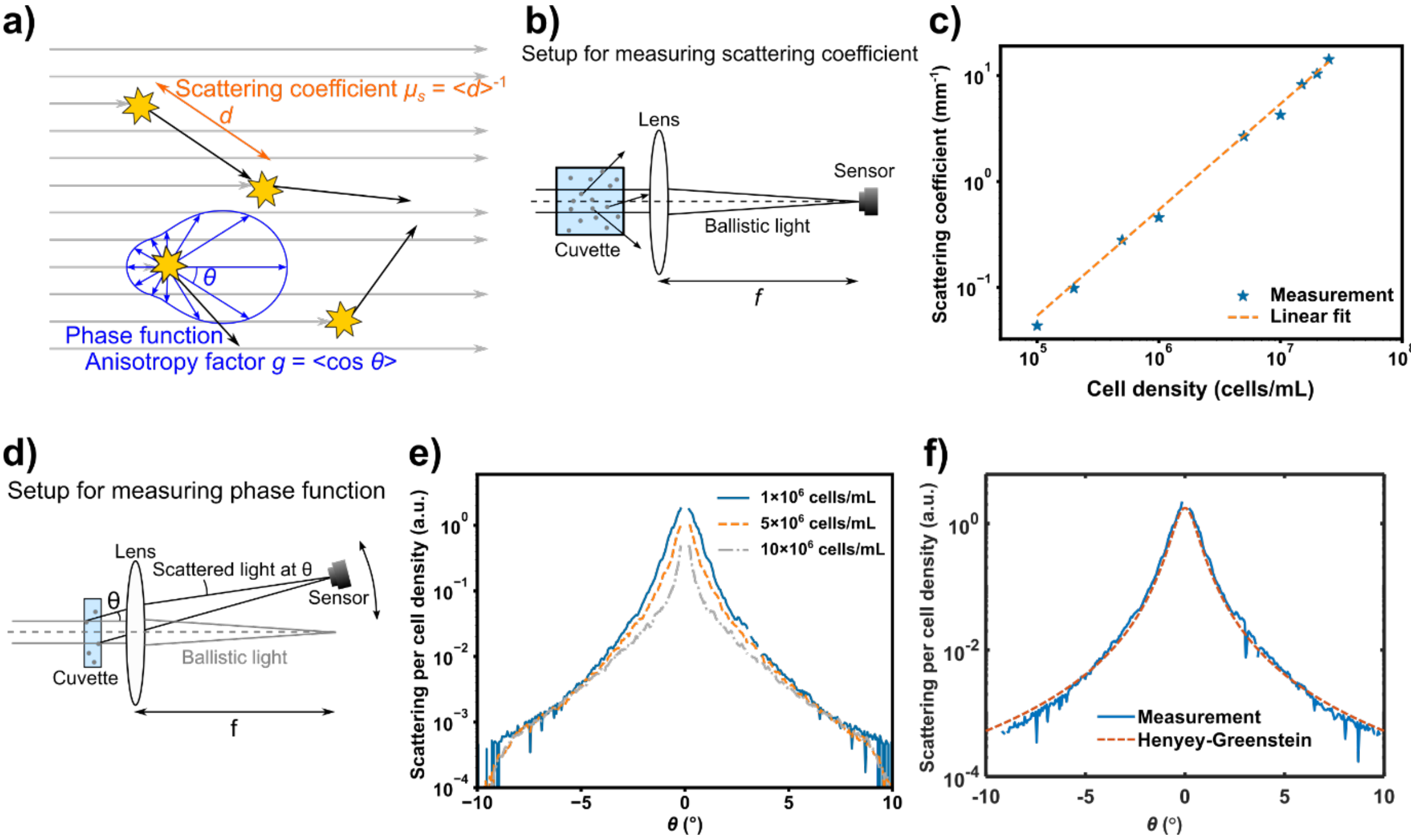

**Figure 2.** Characterization of scattering properties of cell-loaded hydrogels. **a.** Schematic of scattering properties of a medium. Yellow stars represent scatterers in the medium. **b.** Schematic figure of the setup for measuring the scattering coefficient. **c.** Measured scattering coefficient versus the human fibroblast density. **d.** Schematic setup for measuring the phase function. **e.** Measured scattering signal per cell density at various fibroblast concentrations. **f.** Measured scattering signal is fitted with the Henyey-Greenstein phase function to determine the anisotropic factor *g*. The signal is fitted to the curve gradient rather than the absolute value.

The phase function is obtained using the same setup by rotating the sensor around the sample to measure the light distribution at various angles $\theta$, as illustrated in Figure 2d. Then, the scattering signal is obtained by subtracting the signal of the control group (without cells) from the cell-loaded GelMA sample to eliminate the background noise. Figure 2e shows the scattering signals measured at various cell densities and normalized by the cell density. In the single scattering region, where the scattered light is collected after one scattering event, the scattering signal is proportional to the scatterer concentration, and its shape is determined by the phase function. Therefore, the scattering signal per cell density should be independent of the cell density. As the cell density increases, multiple scattering events may occur before the signal is collected by the detector, resulting in a scattered light distribution that is different from that of single scattering in shape. As shown in Figure 2e, signals per cell density at smaller angles are lower in a more concentrated cell suspension (5 and $10\times10^6$ cells $mL^{-1}$), indicating the existence of multiple scattering events: photons with small scattered angles experience another scattering event and shift to larger angles. Therefore, $1\times10^6$ cells $mL^{-1}$ is a suitable concentration for phase function measurement, ensuring enough scattering signals without the inclusion of multiple scattering. The measured angle ranges from 0.15° to 10°. The lower limit is determined by the focused transmitted beam diameter, and the upper limit by the noise level of the detector. A higher scattering signal can be achieved by increasing the incident intensity, at the expense of heating up the sample. Because of the angle range of the measurement, the whole phase function (from 0° to 180°) cannot be obtained on this setup. Previous studies have shown that the Henyey-Greenstein phase function provides a good estimation for the phase function of cell suspensions[22], [23]. Therefore, the scattering signal is fitted to the Henyey-Greenstein function to obtain a phase function estimation, as shown in Figure 2f. The signal curve is fitted by the gradient instead of the absolute value because the scattering signal is not normalized, as in the definition of the phase function, due to the limited angle range. The fitted anisotropic factor $g$ for the tested human fibroblasts is around 0.986. The fitted Henyey-Greenstein phase function and the measured scattering coefficient are used as parameters by our ray-tracing simulation to model TVAM in a scattering environment.

### 2.2. Printing fidelity improvement with scattering correction

Our physics-based inverse rendering algorithm models different physical effects, such as vial geometry, absorption, scattering, and chemical diffusion[5], [19], [20], [24]. It enables robust pattern optimization for various printing scenarios and also acts as a simulation environment for testing the printing fidelity and feasibility of different projected patterns. In this study, the scattering effect of different scattering coefficients (cell densities) on the printing fidelity is thoroughly studied using this tool. A biomimetic vascular structure (Figure 3a), generated with a recently reported algorithm[25], is used for testing the printing fidelity as well as whether the simulation truly reflects the experimental results. This model (4 mm in diameter and 6 mm in height) represents a relevant platform for tissue-mimicking constructs and integration into organ-on-chip systems. Importantly, its dimensions also present a significant challenge for TVAM when using scattering resins, as negative features such as microchannels are particularly susceptible to light-dose blurring caused by scattering. The channels span from the edge of the object to its center, where the ballistic light is minimal according to the Beer-Lambert law, posing a high demand on the precise dose distribution in the model. Successful fabrication of such model is indicative of the printability of other constructs under the same scattering conditions.

Simulated light dose at around $2\times10^7$ fibroblasts $mL^{-1}$ (scattering coefficient: 10 $mm^{-1}$, anisotropic factor: 0.986) is shown to illustrate the scattering and the efficacy of the scattering correction. Compared with the target dose (Figure 3b), the simulated results from the uncorrected pattern set fail to reveal the channels in the construct (Figure 3c), indicating that these negative features are not resolvable. Additionally, the light dose at the bottom of the structure is much lower than that at the middle, indicating that the printed object is shorter than the model, with features at the top and bottom parts missing. In contrast to the uncorrected group, light-dose slices from the scattering-corrected pattern set still exhibit channel structures, although the channels are slightly thicker (around 530 μm) than those in the target and exhibit a blurred edge, as shown in Figure 3d. The dose uniformity along the rotation axis ($z$-axis) is also preserved despite the strongly scattering environment, indicating that the whole 3D structure can be attained and maintained perfusable with scattering correction.

To validate the simulation, we adopted silica bead-loaded hydrogels as a phantom to mimic scattering behaviors in the bioprinting scenario. Silica beads (diameter 3.15 μm) dispersed in 7% GelMA hydrogels show highly forward scattering (fitted anisotropic factor: 0.985) and can match the scattering coefficient of different cell densities by tuning the particle concentration. In this study, the scattering properties of 0.40 wt% silica beads match those of $2\times10^7$ fibroblasts $mL^{-1}$. Photos of printed constructs at various particle concentrations using pattern sets without or with scattering correction are shown in Figure 3e. At 0.24 wt% particle concentration (corresponding to ~ $1.2\times10^7$ fibroblasts $mL^{-1}$), printing without scattering correction barely results in perfusable channels, serving as the printing limit of the pattern optimized without scattering correction. At a concentration of 0.32 wt% (corresponding to ~ $1.6\times10^7$ fibroblasts $mL^{-1}$), one channel is leaking while another one is not perfusable, showing that overpolymerization and underpolymerization occur in the same construct, making it impossible to print such a vascular model by fine-tuning the projected light dose. When increasing the particle concentration to 0.48 wt% (corresponding to $2.4\times10^{7}$ fibroblasts $mL^{-1}$), no sign of channel features can be seen within the print, and the printed object is slightly shorter than the designed height. With scattering correction, however, all the prints are perfusable and preserve their height compared to the uncorrected groups. As a rule of thumb, this screening also shows that the scatterer concentration limit with scattering correction roughly doubles that without scattering correction. Printing experiments with silica beads as the phantom confirm the efficacy of scattering correction as well as the simulation as a powerful tool for predicting the printing outcome.

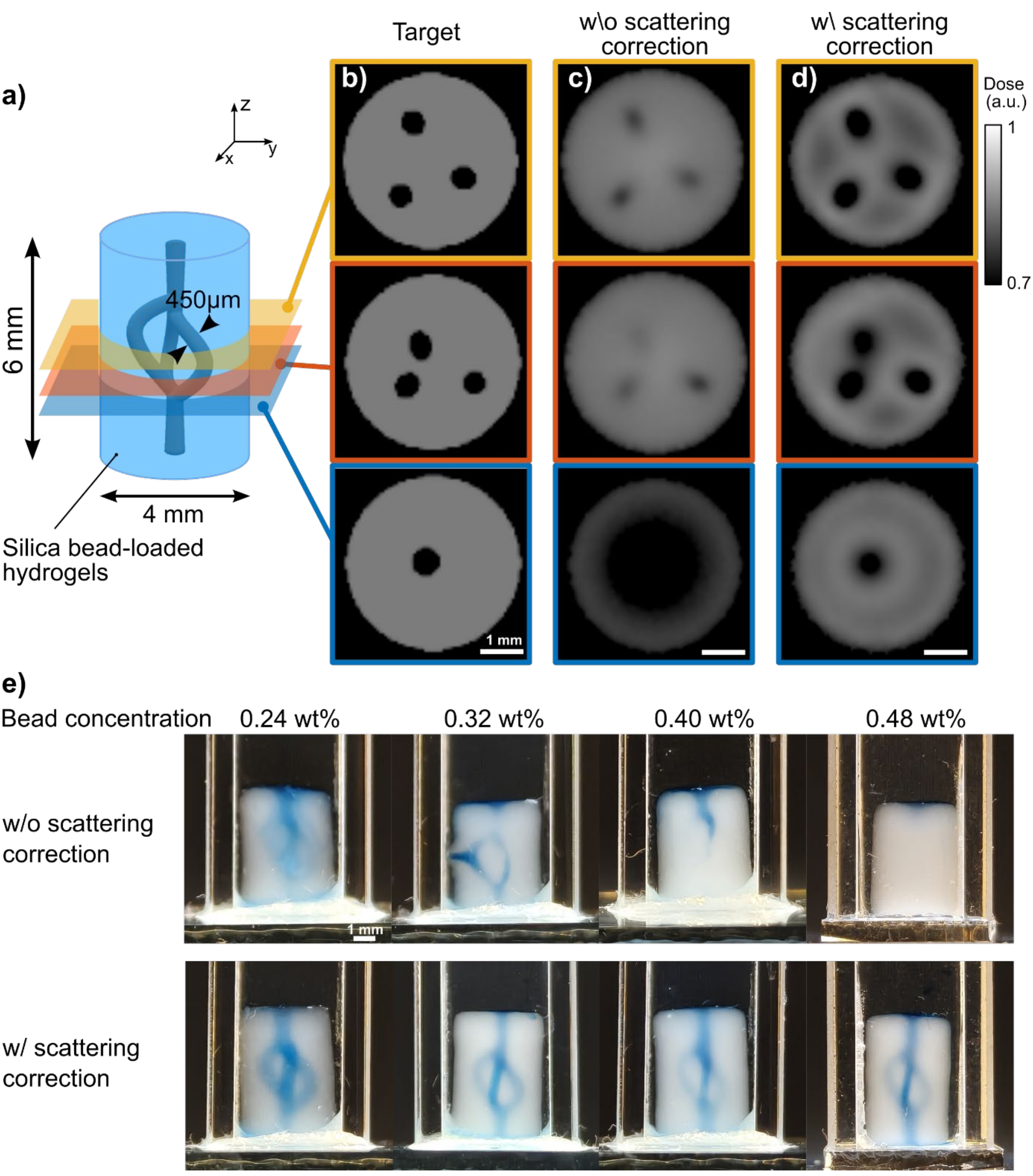

**Figure 3.** Printing fidelity improvement of scattering correction, demonstrated in silica bead-loaded hydrogels. **a.** 3D model of a biomimetic vascular construct. Channel diameter down to 450 µm. Slices of the targeted dose (**b**), simulated light dose of the uncorrected pattern set (**c**), and the scattering-corrected pattern set used in a scattering environment (**d**), each at three different *z*-positions to illustrate the 3D dose distribution. a.u., arbitrary units. **e.** Photos of printed constructs at various particle concentrations using pattern sets without or with scattering correction. Channels are perfused with a blue dextran solution. The particle concentration of printing limit using patterns with scattering correction is around twice that of printing without scattering correction. Scale bars: 1 mm.

## 2.3. Bioprinting with scattering correction

After confirming the efficacy of the computational scattering correction approach both *in silico* and experimentally using silica bead phantoms (Figure 3), we extended this method to the fabrication of both negative and positive features at record-high cell densities of $2\times10^7$ human fibroblasts $mL^{-1}$. To demonstrate this, we employed a photoclick gelatin-based resin with optical properties similar to GelMA but faster crosslinking kinetics (5% gelatin thiol–norbornene, Gel-SH/NB). At such high cell densities, the cellular volume fraction reduces network stability, making rapid formation of mechanically robust hydrogels desirable. In this context, attempts to print and perfuse GelMA hydrogels failed due to insufficient structural stability.

As shown in Figure 4a-i, the use of scattering-corrected patterns enabled successful fabrication of a vascular model in a highly turbid bioresin containing $2\times10^7$ cells $mL^{-1}$. The successful printing of negative features (channels) was further verified by confocal microscopy in combination with perfusion of 2 MDa FITC–dextran (Figure 4a-ii-iii). Moreover, we demonstrate the feasibility of printing positive features at similarly high cell densities using a bear-shaped model. As shown in Figure 4b-ii, simulations predict the recovery of fine positive features (i.e., bear ears) when applying scattering corrections, in agreement with experimental results (Figure 4b iii–iv).

Notably, these experiments were performed using fixed cells. Attempts to print with live cells at concentrations above 1-$1.5\times10^7$ cells $mL^{-1}$ failed due to oxygen consumption by cellular respiration. Oxygen plays a central role as the primary inhibitor in free-radical polymerization and is therefore essential for establishing the chemical contrast required in TVAM printing[24]. At high cell densities, particularly for cell types with high oxygen consumption rates (OCR)[26], oxygen depletion can occur within minutes after bioresin preparation. This continuously alters the effective light dose required for pattern formation, eventually reaching a regime in which oxygen is nearly exhausted and chemical contrast is lost, resulting in uniform bulk gelation rather than spatially resolved structures.

In this work, which focuses on computational correction of cell-induced scattering, the oxygen consumption issue is circumvented by rapidly fixing the cells prior to printing. Nevertheless, oxygen depletion remains an open challenge that must be addressed to enable downstream applications of high-cell-density TVAM bioprinting. Notably, the severity of this limitation is expected to vary with the cellular composition of the construct, as metabolically active cells such as cardiomyocytes and hepatocytes exhibit substantially higher oxygen consumption rates than less oxidative cell types, such as chondrocytes and many stem cell populations[26]. Nevertheless, our scattering correction approach applies to all cell types when the corresponding scattering properties are measured.

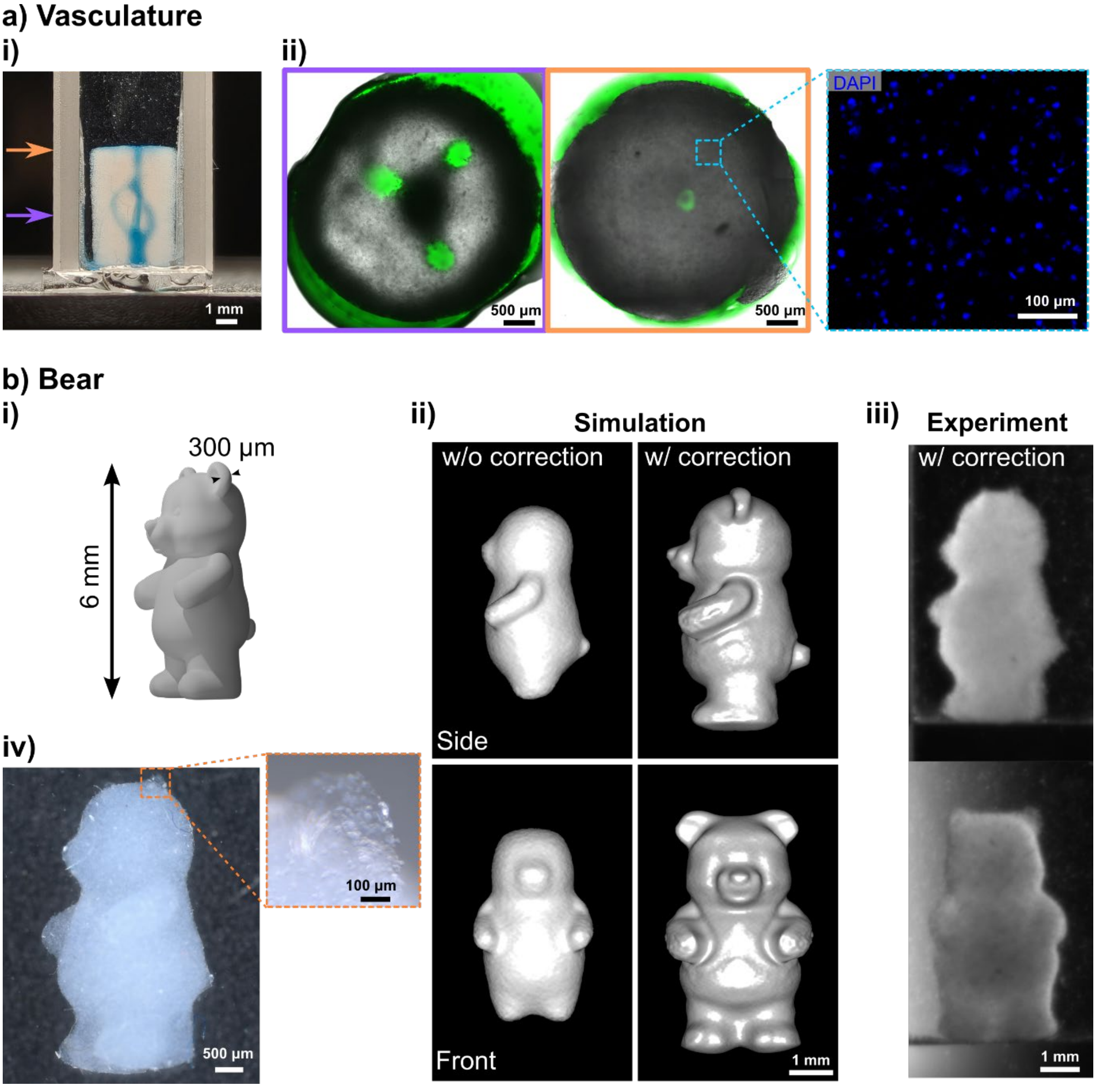

**Figure 4.** Bioprinting at $2\times10^7$ cells/mL with scattering correction. **a.** Demonstration of printing with a vascular-like model. Picture of the printed construct perfused with blue dextran (i). Confocal images of the top and middle cross-section of the cell-laden construct perfused with 2 MDa FITC-dextran (green), and details of cell nuclei (DAPI staining, blue) (ii). **b.** Demonstration of positive feature printing with a bear model (i). Isosurface of the simulated light dose shows the recovered positive features with scattering correction (ii). Isosurface threshold: 0.67. The simulation results are supported by the printed construct shown in the picture (iii) and the microscopic images (iv).

## 2.4. Compatibility with refractive-index engineering

In recent years, partial mitigation of cell-induced scattering in light-based printing has been explored through the use of high refractive-index matching agents such as iodixanol (IDX), without relying on complex computational corrections[13], [27]. In these approaches, scattering is reduced by tuning the refractive index of the resin to minimize the mismatch between cellular and extracellular medium refractive indices. In this work, we demonstrate that this material-based strategy can be further leveraged to increase the upper limit of printable cell density. Moreover, we extend our study to a heterogeneous cell system comprising human foreskin fibroblasts (HFFs) and human pancreatic ductal epithelial cells (HPDE). Interestingly, we find that HPDEs exhibit a very similar anisotropic factor ($g$ = 0.985) compared to HFFs ($g$ = 0.986), but a much lower scattering coefficient ($\mu_s$= 5.5 mm$^{-1}$ at $2\times10^7$ cells mL$^{-1}$) compared to HFFs ($\mu_s$= 10.8 mm$^{-1}$ at $2\times10^7$ cells mL$^{-1}$), further highlighting cell-type-dependent optical properties. These measurements highlight the importance of experimental determination of these values for a physically accurate computational scattering correction.

We verified by direct measurement that a proper concentration of IDX can reduce the scattering effect of cell-laden bioresin. The refractive index of the bioresin linearly increases with the IDX concentration, as shown in Figure 5a. This increase reduces the refractive index mismatch between the bioresin and the embedded cells, resulting in a steady decrease of the scattering coefficient (Figure 5b). Interestingly, adding IDX has a minimal effect on the anisotropic factor (Figure 5c). These results show that a high concentration of IDX reduces the scattering effect by substantially decreasing the effective scattering cross-section of the embedded cells, instead of altering the scattering directionality.

Finally, by integrating material tuning (30% IDX) with computational scattering compensation, we demonstrate successful fabrication of a perfusable vascular model (Figure 5d–f) at an unprecedented cell density of $4.1\times10^7$ cells mL$^{-1}$ ($1.95\times10^7$ cells mL$^{-1}$ of HFFs and $2.19\times10^7$ cells mL$^{-1}$ of HPDEs). 10% Gel-SH/NB is used to enhance the network stability at high cellular volume fraction and high IDX volume fraction. Altogether, these results expand the operational window of TVAM bioprinting, which was previously considered limited to highly transparent and low-cell-density resins, and open new avenues toward the fabrication of physiologically relevant high-cell-density tissues.

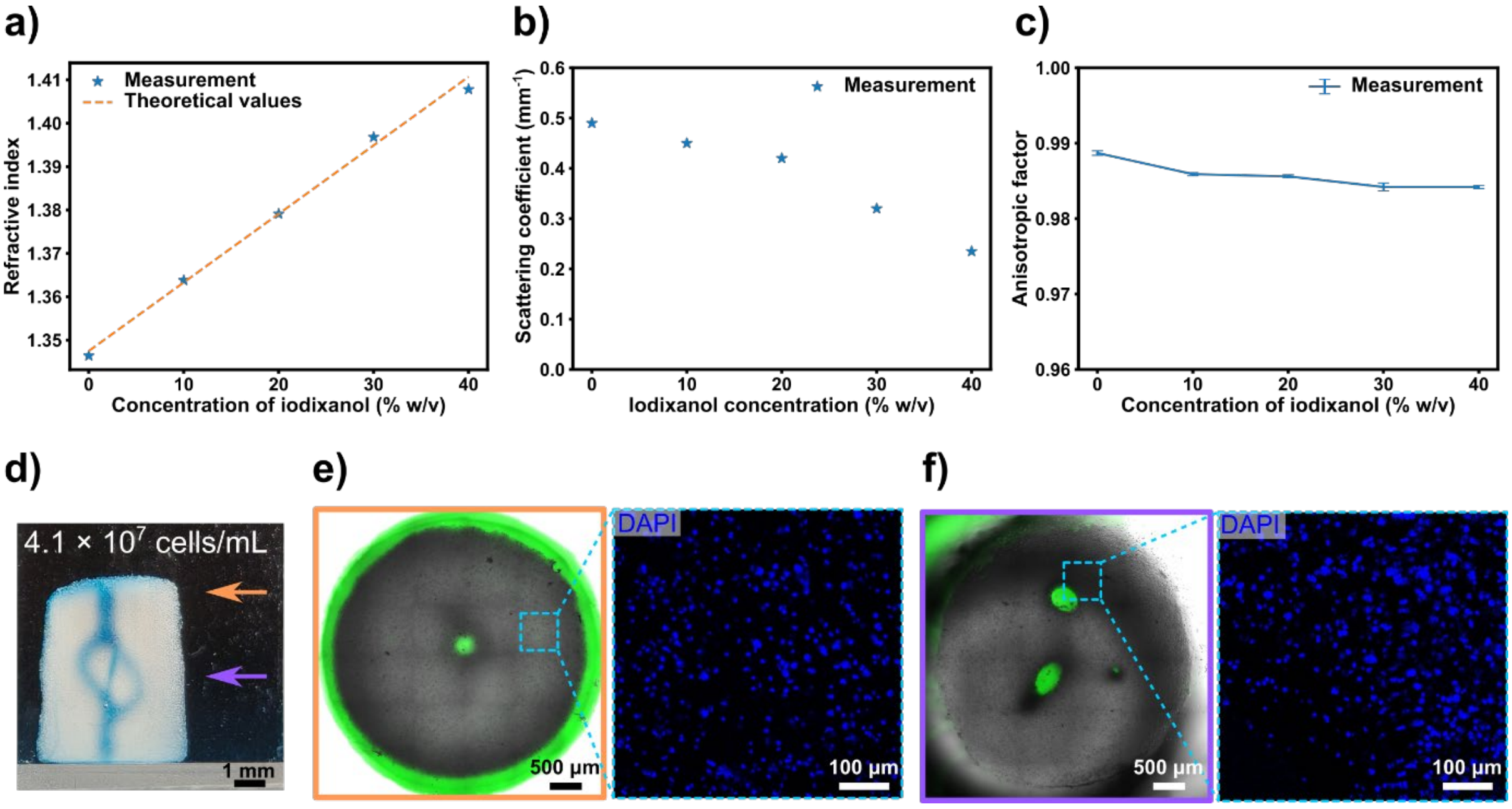

**Figure 5.** Bioprinting at 4.1×10$^7$ cells/mL with scattering correction and refractive index engineering. **a.** Refractive index of 7% GelMA bioresin with various IDX concentrations. The theoretical values assume linear refractive index increments for aqueous mixtures. Scattering coefficient (**b**) and anisotropic factor (**c**) of 10$^6$ fibroblasts mL$^{-1}$ cell-loaded bioresin with various IDX concentrations. **d.** Picture of the printed construct perfused with blue dextran. Confocal images of the top (**e**) and middle (**f**) cross-section of the cell-laden construct perfused with 2 MDa FITC-dextran (green), and details of cell nuclei (DAPI staining, blue).

## 3. Discussion

This study demonstrates a practical, physics-based pattern optimization to improve tomographic volumetric printing fidelity in cell-laden hydrogels by optical characterization and ray-tracing-based projection optimization. The experiments and simulations show the cell density limit for tomographic volumetric bioprinting without scattering correction. When the cell density is higher than the limit, the scattering correction is crucial for realizing features (especially negative features) and maintaining the functions of the printed construct. One of the effects of scattering on the construct is that light doses at the top and bottom of the structure are severely attenuated compared to those in the middle. By incorporating parameters such as vial optics, absorption, and scattering into an inverse rendering model, our framework detects this effect and optimizes the pattern set accordingly, resulting in a relatively uniform light dose along the whole structure.

The computational method is not exclusive to material strategies. This study has demonstrated its compatibility with the refractive index engineering approach. Other scattering-mitigating techniques such as wavelength-shifting (using upconversion nanoparticles[28], [29], [30], triplet-triplet annihilation[31], or red/near-infrared initiator system[32]) can be used in concert with pattern optimization to extend the printable cell density of tomographic volumetric bioprinting towards physiologically relevant cell densities.

Beyond compensating for scattering in one bioink, this computational framework can also accommodate heterogeneous or layered bioinks with different cell types or concentrations by designing a simulation model accordingly. It is also compatible with hybrid fabrication workflows (e.g., TVAM integrated with extrusion-based printing[12], [33], overprinting[12], [20]) and can also be explored in other light-based 3D printing techniques such as xolography[34].

The principal limitation of this scattering correction method is its computational budget: high-fidelity differentiable ray tracing with many projection angles and fine voxel grids is expensive in time and memory and can make large or high-throughput designs onerous. In practice, we mitigate this with several engineering options: use a larger but coarser exploration grid during early parameter testing (to rapidly scan design hyperparameters and convergence behavior) and then check promising candidates at higher resolution; define the active projector area and the sensor volume to be only slightly larger than the model. These strategies trade extra design iterations for a much lower computational cost than directly optimizing at full experimental resolution from scratch.

Notably, although this work successfully demonstrates the integration of physics-based scattering correction in TVAM and identifies its practical upper limit, a major challenge for biomedical applications remains cellular respiration and the resulting oxygen depletion at cell concentrations above approximately 1-$1.5\times10^{7}$ cells $mL^{-1}$. Because oxygen is the primary inhibitor of free-radical polymerization, its consumption during the printing process (typically 10–60 min) continuously alters the required light dose. At sufficiently high cell densities, oxygen depletion can become so extensive that the inhibition required to generate the chemical contrast underlying TVAM is nearly eliminated. Consequently, extending this approach to higher cell densities (2–4×10$^{7}$ cells mL$^{-1}$) will require additional strategies, such as incorporating inhibitors that are not consumed by cells (e.g., TEMPO[35] or DMPO[36]), transiently reducing cellular respiration through chemical or metabolic modulation, or combining these approaches. Alternatively, photoinitiator and radical-free systems, such as those based on photouncaging[37], [38], and photolysis[39], [40], could be adopted to reduce the influence of oxygen inhibition.

Another important factor is the vial diameter. For a defined cell density and resin composition, increasing the vial diameter increases the optical path length through the scattering medium, resulting in greater light scattering and image blurring. Consequently, the maximum printable cell density decreases as the vial diameter increases. Specifically, doubling the vial diameter reduces the maximum printable cell density by a factor of 2. In this work, 5 mm inner diameter vials enabled successful printing at cell densities up to 2×10$^{7}$ human fibroblasts mL$^{-1}$ or 4×10$^{7}$ human pancreatic ductal epithelial cells mL$^{-1}$ because of their lower scattering coefficient. Increasing the vial diameter to 10 mm would reduce the maximum printable cell density to approximately 1×10$^{7}$ human fibroblasts mL$^{-1}$, assuming all other printing parameters remain unchanged. Constructs with larger features or positive features of the same size can be printed at higher cell densities.

Further improvement can be made on the scattering measurement. Accurate scattering parameters are essential but do not require a separate goniometry setup: given that the phase function is similar for various cell types and different refractive indices of the bioresin, the scattering coefficient can be obtained through the transmission measurement, which can be integrated into the TVAM setup. This *in-situ* calibration closes the loop between characterization and optimization while remaining non-intrusive and practical for routine use.

## 4. Conclusion

In summary, the physics-based, ray-tracing algorithm for pattern optimization provides a general and experimentally validated approach to mitigate many scattering-induced failures in tomographic volumetric bioprinting. Its effectiveness has been tested in high-cell-density bioresins, where it is challenging to print functional constructs using TVAM. By significantly extending the cell density range of TVAM, these findings open new avenues for the rapid fabrication of centimeter-scale, high-cell-density constructs that are potentially physiologically and functionally closer to native tissues. Furthermore, its flexibility with other scattering mitigation approaches makes it applicable to a wide range of biofabrication scenarios and hybrid workflows.

## 5. Experimental Section

### Synthesis of Gelatin-Methacryloyl (Gel-MA)

10 g type A porcine gelatin powder (Sigma, G2500) was fully dissolved at 10% w/v in phosphate buffered saline (PBS) 1× at 55 °C. 6 mL of methacrylic anhydride (Sigma, 760-93-0) was added dropwise for gelatin modification at 55 °C for 1 h. The solution was then dialyzed (MWCO: 14 kDa, Roth AG) for 5 days, lyophilized, and stored away from light at -20 °C until use. The degree of substitution was estimated to be ∼0.25 mmol of MA per gram of gelatin with $^{1}$H-NMR in $D_2O$ using internal standard 3-(trimethylsilyl)-1-propanesulfonic acid (DSS).

### Synthesis of Gelatin-Norbornene (Gel-NB)

Gel-NB was synthesized as previously described by Rizzo *et al.*[41] In short, 20 g of gelatin type A (Sigma, G2500) from porcine skin was dissolved in 0.5 M pH 9 carbonate–bicarbonate buffer at 10% at 40 °C. When completely dissolved, 0.4 g of cis-5-norbornene-endo-2,3-dicarboxylic anhydride (CA, Chemie Brunschwig AG) was added to the reaction mixture under vigorous stirring. 0.4 g of CA was added every 10 min for a total amount of 2 g. 20 minutes after the last addition, the solution was diluted with 200 mL of prewarmed mQ $H_2O$. Next, 2 g of NaCl was added, and the solution was filter-sterilized (0.2 µm) and dialyzed (MWCO: 14 kDa, Roth AG) for 4–5 days against mQ $H_2O$ at 30 °C with frequent water changes before freeze-drying. Lyophilized Gel-NB was stored at -20 °C before use. The degree of substitution was estimated to be ∼0.17 mmol of NB per gram of gelatin with $^{1}$H-NMR in $D_2O$ using the internal standard DSS.

### Synthesis of Gelatin-Thiol (Gel-SH)

Gel-SH was synthesized as previously described by Rizzo *et al.*[42] In short, 10 g of type A porcine gelatin (Sigma, G2500) was dissolved in 500 mL of 150 mM pH 4.5 MES buffer. Then, 0.48 g of 3,3'-dithiobis(propionohydrazide) (DTPHY, Chemie Brunschwig AG) was added to the solution. Once dissolved, 1.5 g of 1-ethyl-3-(3-dimethylaminopropyl)carbodiimide (EDC, Roth AG) was added to the reaction mixture. After 24 h, 3.3 g of reducing agent tris(2-carboxyethyl)phosphine (TCEP, Chemie Brunschwig AG) was added to the mixture, and reduction was left to proceed for 6 h. Then, 1 g of NaCl was added to the solution, which was then dialysed (MWCO: 14 kDa, Roth AG) against acidified (pH 4) mQ $H_2O$ for 4-5 days with frequent water changes. The Gel-SH solution was then sterile filtered before freeze-drying. Lyophilized Gel-SH was stored at -20 °C before use. The degree of substitution was estimated to be ~ 0.26 mmol of SH per gram of gelatin with $^1$H-NMR in $D_2O$ using the internal standard DSS.

**Cell culture**

Human foreskin fibroblasts (HFF-1, SCRC-1041 ATCC) and wild-type human pancreatic ductal cells (HPDE-wt) were cultured in DMEM without phenol red, supplemented with 1% Penicillin-Streptomycin (Gibco), 2% L-glutamine (Gibco), and 10% fetal bovine serum (FBS) (Gibco). Cells were maintained in a humidified $CO_2$ incubator at 37 °C and 5% $CO_2$. The media was changed every other day, and cells were used at passage 7-13.

**Bioresin preparation**

Gel-MA and Gel-SH/NB (1:1 SH/NB ratio) resins were prepared by dissolving the lyophilized material at the desired concentration in DMEM without phenol red, with LAP at a concentration of 0.5 mg mL$^{-1,}$ and filter-sterilized at 37 °C. For bioresin with IDX, Gel-MA or Gel-NB stock solution was mixed with IDX solution (OptiPrep, Sigma) at the desired concentration. Trypsinized cells were counted, spun down, and resuspended in a minimum volume of PBS with 0.5 mg mL$^{-1}$ LAP. Cell suspension was then mixed with photoresin stock solutions (10% Gel-MA or 12% Gel-SH/NB) to result in 5%, 7%, or 10% bioresin.

**Silica beads-loaded resin preparation**

Silica bead solution (d = 3.15 μm, Spherotech) was vortexed to form a homogeneous dispersion and mixed with photoresin stock solution (10% GelMA) and LAP to result in 7% GelMA, 0.5 mg mL$^{-1}$ LAP, and various silica bead concentrations. The resin is distributed into vials and dipped into ice to prevent bead sedimentation.

**Scattering characterization**

The scattering coefficient and the phase functions are measured on the same goniometric setup. Light from a 532 nm laser (Coherent, Verdi V10) propagates through a tilted optical window (Thorlabs), a beam chopper at 50 Hz, and the scattering hydrogels encapsulated in a cuvette. A reference detector (Thorlabs, PDA36A-EC) collects the reflected light from the tilted optical window to compensate for the laser power fluctuation. The beam chopper is applied for the background correction. Then the light signal after the sample is focused by a lens (f = 250 mm) and collected by the detector (Thorlabs, PDA36A-EC). The goniometer is rotated around the sample with an arm of 300 mm. A tube and a 1-mm-diameter pinhole are placed in front of the detector to ensure small scattering angle detection. The cuvette for phase function measurement is a demountable cuvette with a light path of 0.5 mm (FireflySci, Type 20) to achieve single scattering within the sample.

Fibroblasts were detached, counted, centrifuged, and resuspended at various cell densities into 7% GelMA hydrogel. The scattering properties of the silica beads are measured with the silica beads dispersed at various bead concentrations in GelMA hydrogel. The solution was transferred into the cuvette, which was then placed at 4 °C to allow GelMA physical gelation. The measurement was conducted at low laser power (10 mW) to avoid gel melting and sedimentation of suspended cells.

**Pattern optimization and light dose simulation**

The optimization for producing the required tomographic patterns was performed using the algorithm described in a previous work[19]. Scattering correction was applied by incorporating the experimentally measured scattering coefficient and anisotropic factor of the cell-loaded hydrogels into the Monte-Carlo ray-tracing model used as the forward simulator. In this model, the scattering coefficient determines the mean free path between successive scattering events. Absorption $\mu_a$ and scattering coefficient $\mu_s$ are jointly represented through the extinction coefficient $\mu_{ex}$ and optical albedo $a$:

$$\begin{cases} \mu_{ex}=\mu_a+\mu_s \\ a=\dfrac{\mu_s}{\mu_{ex}}=\dfrac{\mu_s}{\mu_s+\mu_a} \end{cases}$$

At each scattering event, the change in photon direction is sampled from the Henyey–Greenstein phase function parameterized by the measured anisotropic factor $g$. Together, these parameters define the ray-tracing model used by the differentiable projector to compute the light-dose distribution inside the opaque bioresin.

The lower and upper dose thresholds were set to 0.6 and 0.9, respectively, and the remaining parameters were matched to the measured optical properties. All calculations were performed on an NVIDIA L40S GPU. For scattering-corrected patterns, parameters such as the number of light paths generated per projection pixel and the maximum number of scattering events allowed before terminating a path are set based on the scattering coefficient. For each print, we used a set of 300 grayscale tomographic patterns with an angular interval of 1.2°, and the cylindrical vials were rotated at a constant angular speed of 105°/s.

**Printing process**

Cell-loaded resin was prepared as previously described and poured into ethanol-sterilized glass test tubes (inner diameter 5 mm) with a hermetically sealing cap. All these operations were conducted under sterile conditions in a biosafety cabinet. For resin containing $\geq 2\times10^7$ cells $mL^{-1}$, cells were fixed to avoid extensive oxygen consumption during the printing session. In short, trypsinized cells were mixed with 4% PFA in PBS and centrifuged. The cell suspension was then washed in PBS and centrifuged twice to remove residual PFA.

Once filled, the glass vials were readily dipped into ice to gel the GelMA. They were mounted onto a custom tomographic volumetric setup for printing. In this printer, UV light from a 400 nm laser diode (Civil Laser) is expanded and projected on a digital micromirror device (DMD, pixel size = 13.7 µm, Vialux, VIS-7001), which displays the tomographic patterns. Two plano-convex lenses project the images from the DMD onto the rotating vial with the final pixel size of 20.45 µm. The vial is set to rotate using a high-precision stage (Zaber, X-RSW60C). Prints were completed in less than 10 seconds.

After printing, glass vials were heated by dipping them into a water bath at 37 °C. Warm water was gently pipetted into the glass tubes, then they were gently manually agitated to rinse away the unpolymerized GelMA. Blue dextran was injected into channels of rinsed bioprinted fibroblast-laden constructs to reveal the inner structure.

**Imaging**

Images of the printed constructs were obtained with a phone or a Fujifilm XT-3 camera. Details of the bear print (Figure 4b-iv) were obtained with a Keyence digital microscope (VHX-5000). Confocal images were obtained with a confocal inverted microscope (Leica, SP8).

**Acknowledgement**

The authors thank Prof. Chiara Tonda-Turo from Politecnico di Torino for providing HPDE-wt. This work was supported by the Swiss National Science Foundation under grant number 10007068 “Neural precision Holographic Volumetric Additive manufacturing” (F.W.) and the Swiss National Science Foundation Return CH Postdoc.Mobility fellowship, P5R5-3_235066 (R.R.).

**Data Availability Statement**

The data that support the findings of this study will be made publicly available upon manuscript acceptance. Any remaining supporting data are available from the corresponding author upon reasonable request.

## Conflict of interest

Christophe Moser is a shareholder of Readily3D SA. Qianyi Zhang is currently an employee at Readily3D SA. The research presented in this manuscript was conducted entirely during her prior affiliation with EPFL. The current employer was not involved in the study design, data interpretation, or manuscript preparation. All the other authors declare no conflicts of interest.